\documentclass{ws-ijmpcs}

\usepackage[nice,loose]{units}
\usepackage{amsmath}
\usepackage{graphics}

\begin{document}

\markboth{Mauro Mariani, Milva G. Orsaria \& H\'ector Vucetich}
{Simplified Thermal Evolution of Proto-hybrid Stars}

%
\catchline{}{}{}{}{}
%

\title{Simplified Thermal Evolution of Proto-hybrid Stars}

\author{Mauro Mariani$^{a,b,\dag}$,
M. Orsaria$^{a,b, c\ddag}$ and
H. Vucetich$^{b\S}$}

\address{
$^{a}$ Grupo de Gravitaci\'on, Astrof\'isica y Cosmolog\'ia,\\
  Facultad de Ciencias Astron{\'o}micas y Geof{\'i}sicas, Universidad Nacional de La Plata,\\
  Paseo del Bosque S/N (1900), La Plata, Argentina.\\
$^{b}$ CONICET, Godoy Cruz 2290, 1425 Buenos Aires, Argentina.\\
$^{c}$ Department of Physics, San Diego State University, 5500 Campanile Drive,\\
  San Diego, CA 92182, USA.\\
$^{\dag}$mmariani@fcaglp.unlp.edu.ar, $^{\ddag}$morsaria@fcaglp.unlp.edu.ar,\\
$^{\S}$vucetich@fcaglp.unlp.edu.ar
}

\maketitle

\begin{history}
\received{Day Month Year}
\revised{Day Month Year}
\published{Day Month Year}
\end{history}

\begin{abstract}
We study the possibility of a hadron-quark phase transition in the interior of neutron stars, taking into account different schematic evolutionary stages at finite temperature. Furthermore, we analyze the astrophysical properties of hot and cold hybrid stars, considering the constraint on maximum mass given by the pulsars  J1614-2230 and  J1614-2230. We obtain cold hybrid stars with maximum masses $\geq 2$ M$_{\odot}$. Our study also suggest that during the proto-hybrid star evolution a late phase transition between hadronic matter and quark matter could occur, in contrast with previous studies of proto-neutron stars.
\end{abstract}

\ccode{PACS numbers:}

\section{Introduction}	

A proto-neutron star (proto-NS) is the remaining old degenerated core of an intermediate mass star, between 10 and 25 solar masses at the zero age main sequence, after a supernova type-II or type-Ib/Ic explosion. Its evolution could be described roughly by three successive isentropic stages after which the resulting object is a cold and stable NS.\cite{Burrows:1986,Steiner:2000}

The inner and outer cores of NS contain most of its mass but its equation of state (EoS) has not yet been well determined and it has been suggested that a phase transition from hadronic to quark matter could occur, giving rise to a hybrid star (HS), a neutron with a quark matter core.

The discovery of the massives PSR J1614-2230 and PSR J1614-2230 ($\sim2$ M$_{\odot}$) neutron stars \cite{Demorest:2010} has placed bounds on a wide variety of EoS. In this work we study hybrid stars, where the EoS is constructed by assuming a sharp phase transition between hadronic matter and quark matter. For the hadronic phase we use  a table with the EoS created for astrophysical simulations of proto-NS \cite{Shen:2010a}. For the quark matter phase we use the Field Correlator Method (FCM), which has been already used to describe quark matter inside cold HS \cite{Logoteta:2013,Burgio:2016}.

\section{Quark matter and isentropic stages of thermal evolution}

\begin{figure}[tb]
  	\centering
  	\includegraphics[width=0.6\columnwidth]{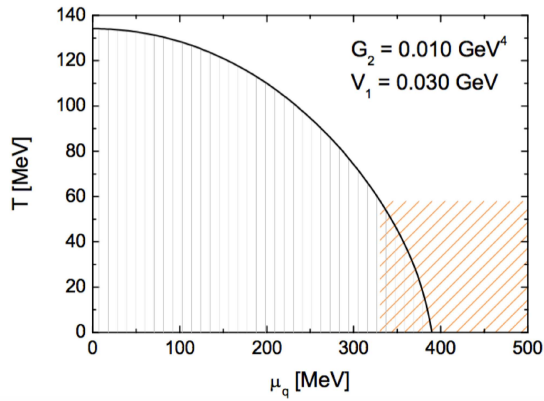}
  	\caption{Schematic phase diagram in the $T-\mu_q$ plane for a representative set of FCM parameters  $V_1$ and $G_2$. The shaded region with vertical lines indicates the confined phase. The diagonal lines indicate the region of the phase diagram where proto-neutron stars are, assuming that their nuclei are sufficiently dense so that the quark matter phase appears.}
  	\label{phase_diag}%
    \end{figure}

The FCM is a non-perturbative approximation of QCD parametrized through the gluonic condensate, $G_2$, and the quark-antiquark static potential for long distances, $V_1$ \cite{Mariani:2016}. In the NS regime of low temperature and high densities, bottom part of the orange shaded region in Fig. \ref{phase_diag}, the behaviour of $V_1$ and $G_2$ is not well known, so we treat them as free parameters, independent of the temperature and/or the baryonic chemical potential.

The quark-gluon plasma pressure within the FCM is given by \cite{Mariani:2016}
   	\begin{equation}
	P_{qg} = ‎\sum_{q=u,d,s} (P_q+P_{\bar{q}}) + P_g - \frac{9}{64} G_2 \, ,
	\label{pqg}
	\end{equation}
where the quark-antiquark pressure is
	\begin{equation}
		\frac{\pi^2}{T^4} (P_q + P_{\bar{q}}) = \phi_\nu (\frac{\mu_q - V_1/2}{T}) + \phi_\nu (\frac{-\mu_q - V_1/2} {T})\, ,	
	\end{equation}
and
	\begin{equation}
	\phi_\nu (a) = \int_0^\infty du \frac{u^4}{\sqrt{u^2+\nu^2}} \frac{1}{\exp{[\sqrt{u^2 + \nu^2} - a]} + 1}\, , 	
	\end{equation}
with $\nu=m_q/T$, where $\mu_q$ and $m_q$ are the chemical potential and the quarks masss, respectively. The gluon pressure is given by

\begin{equation}
  P_g = 16 \, \frac{T^4}{\pi^2}  
     \mathrm{Li}_4\left(e^{-\frac{9}{8}\frac{V_1}{T}}\right) 
        \, ,
\end{equation}
where $\rm{Li}_s[z]$ is the $Jonqui\grave{e}re's$ $function$. To calculate semi-analytically the quark pressure we built a double power series expansion \cite{Masperi:2004}  in terms of $m_q^2 /(u^2 T^2 + m_q^2)^2$ and $(\mu_q - V_1/2)/T$ . 

The thermal evolution of the proto-HS starts with an entropy per baryon of $s = 1$, where trapped neutrinos are subjected to the condition $Y_{L_e} \simeq 0.4$. In the following stage of thermal evolution, after $\sim 15$ seconds, neutrinos diffuse and warm the star being $s = 2$ and $Y_{\nu} = 0$. Finally, after a few minutes, the star cools down, the last stage is reached and $s = 0$ \cite{Burrows:1986,Steiner:2000}. Supposing that the baryonic mass remains constant during the process of thermal evolution, the gravitational mass of the star, which corresponds to the total stellar energy, changes during the evolution of the proto-NS. Then, a study of the evolution in a gravitational mass - baryonic mass plane must be take into account, giving in this way the possibility to know in which stage takes place the phase transition from hadronic matter to quark matter. 

\section{Gravitational and baryonic mass}

\begin{figure}[tb]
  		\centering
  		\includegraphics[width=0.85\columnwidth]{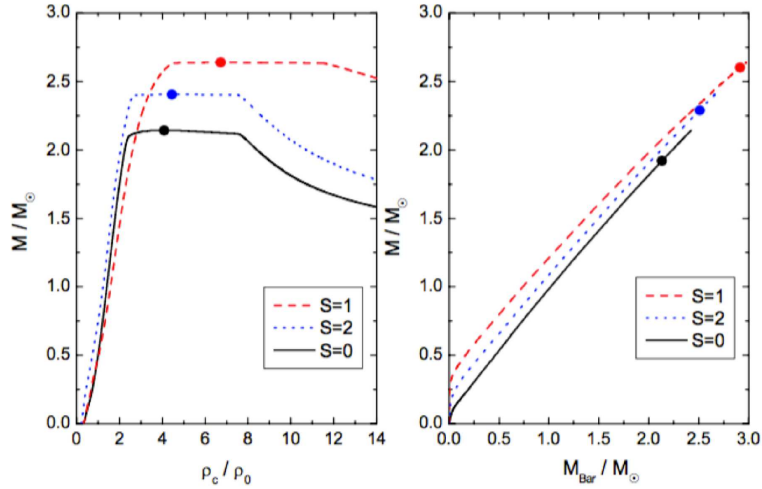}
  		\caption{Gravitational mass versus central density  (left panel) and as a function of the baryonic mass (right panel) for HS obtained with our model. In the right panel, the evolutionary process of the proto-HS can be followed starting from the $s = 1$ curve, downwards along a constant baryonic mass vertical line. Dots indicate the maximum mass star in the left panel and the the appearance of the quark matter core in the right panel, for different stages of the HS thermal evolution. }
      	\label{mbar}%
\end{figure}

Once the hybrid EoS is obtained, we have the input to solve the Tolman-Oppenheimer-Volkoff (TOV) equations, which are the relativistic structure equations of hydrostatic equilibrium and mass conservation for a spherically symmetric configuration. The EoS determines the mass-radius relationship for each group of stars. The gravitational mass as a function of the central density and the baryonic mass are shown in Fig. \ref{mbar}.

 The possibility of a phase transition in the proto-NS is indicated by rounded dots in the $M/M_{\odot}-M_{Bar}/M_{\odot}$ plane, suggesting the appearance of a quark matter core. Configurations with smaller baryonic masses than the associated to the dot in each curve ($M_{Bar}^{(q,\,s=i)}$, $i=1,2,0$) are purely hadronic (left side of the dot), while configurations with greater baryonic masses (right side of the dot) are NS with a quark matter core. From Figs. \ref{mbar}, it can be seen that quark matter could be formed during the constant baryonic mass evolution. Thus, a proto-NS having a baryonic mass $M_{Bar}^{(q,\,s=0)} < M_{Bar} < M_{Bar}^{(f)}$ ending as a stable cold NS, would begin as purely hadronic and become a HS with a quark matter core after the whole cooling process.

\section{Conclusions}

We have obtained stable cold HS that reach 2M$_{\odot}$ for the maximum mass star. According to our results, the pulsars J1614-2230 and J1614- 2230 would contain quark matter in their inner cores. The sequence of snapshots suggests that a quark matter core may be formed during the cooling process between the stages with entropies per baryon $s \simeq 2$ and $s=0$. This is in contrast with the predictions of earlier studies where the deconfinement of quark matter occurs some seconds after the core bounce \cite{Benvenuto:1989qr} or the transition from nuclear to quark matter takes place during the stage with $s \simeq 1$ \cite{Benvenuto:1999}.

Considering the different constant entropy stages, the formation of a quark matter core during the cooling process could be a suggestive possibility, but further analysis on hadronic-quark matter phase transition as well as on thermal evolution of NS should be done before any certainty.


\end{document}